\documentclass[showpacs,preprintnumbers,amsmath,amssymb, superscriptaddress, 12pt ]{revtex4}

\usepackage{graphicx}
\usepackage{amsfonts}
\usepackage{amsmath, amsthm, amssymb}

\usepackage{fancyhdr}

\usepackage[norsk]{babel}
\usepackage[latin1]{inputenc}

\usepackage{amsmath, amsthm, amssymb}

\usepackage{epsfig}	
\usepackage[hang]{subfigure}                
\usepackage[small,bf,hang]{caption}       
\usepackage{ifpdf}   
\usepackage{verbatim}
\ifpdf                                                                  
  \usepackage{ae}                               
  \usepackage{hyperref}                     
\fi

\begin{document}

\newtoks\slashfraction
\slashfraction ={0.7}
\def\slash#1{\setbox0\hbox{$ #1 $}
\setbox0\hbox to \the\slashfraction \wd0{\hss \box0}\not\box0}

\def\dslash {\not{\hbox{\kern-2pt $\partial $}}}
\def\Dslash {\not{\hbox{\kern-4pt $D$}}}
\def\Qslash {\not{\hbox{\kern-3.5 pt $Q$}}}
\def\pslash {\not{\hbox{\kern-2.3 pt $ p$}}}
\def\kslash {\not{\hbox{\kern-2.3 pt $ k$}}}
\def\qslash {\not{\hbox{\kern-1.3 pt $ q$}}}
\def\Aslash {\not{\hbox{\kern-3.0 pt $ A$}}}

\newcommand{\bfk}{{\bf k}}  
\newcommand{\bfx}{{\bf x}}  
\newcommand{\bfp}{{\bf p}}  
\newcommand{\I}{ {\rm{i}} }

\title[]{Analyze of multistage transfer orbit to reach the GEO orbit in a rotational symmetric gravitational field}

\author{S. Ng\footnote{\label{addrNg} Electronic address: ng@stud.ntnu.no}}

\address{Department of Physics, Norwegian University of Technology and Science, 7043 Trondheim, Norway}

\begin{abstract}
Placing satellites in geostationary obital (GEO) is propellant exigeant. It requires large propellant consumption that will take a large part of the total mass. Hence the interest is reduce the mass of propellant loaded both in the rocket and the satellite. Both for improving selection of transfer orbits and propellant. 
\\
\\
In this paper we give a reiew of analysis of launching of satellite to GEO using multistage transfer orbit in order to reduce the mass of propellant. Making this analysis complete, the effect of the general relativity regarding elliptical transfer orbit and its potensial utilization will be briefly reviewed.
\end{abstract}

\pacs{91.10.Sp, 91.10.Op}

\maketitle

\section{INTRODUCTION} 
A circle at an altitude of approxmately $35\,786{\rm km}$ above the equator is known as the geostationary arc. Geostationary orbit, or GEO as it is called, is discovered because of its advantegs. It is situated circular in the equatorial plane, moving tn step with the earth rotation. The geostationary orbit time is close to a solar day which has some advantageous properties. The orbit covers the most of the earth. Moreover, the satellites i GEO are fixed in proportion to each point of the earth surface. Hence, it is possible to construct a system with a single satellite that yields time continuous coverage of an area on the earth. The technology of geostationary orbit is well depeloped and proven today. Other advantegs are there is minimal effect and stable signal strength due to constant ground satellite range in such a orbit. 
\\
\\
However there are disadvantages with a geostationary orbit. The propagation delay is relatively large which affects voice and time sensitive data protocols. And the launch cost is relatively high, since the distance between the geostationary orbit and the earth is large. Using one stage transfer orbit to reach GEO is thus not advantageous due to the need of huge propellant consumption. Besides, satellites in the GEO orbit do not cover the territory around the North and South Pole. About $\pm84^\circ$ the geostationary satellites are no longe visible. This is an inevitable disadvantage. Having satellites in the GEO orbit has been more difficult in the 21th century. The number of GEO satellites has increased significantly. The traffic in the geostationary orbit is large and it is now big crush in the orbit. However, the geostationary orbit is a important but confined resource of communication and science that we always will fight for improving our exploitation. 
\\
\\
Today, due to the large traffic in GEO, the satellite control systems play a vigorous part in communication and space technology. At $19.2^\circ$ 5-ASTRA satellites operate in nominal terms same orbit posistion as GEO when they move in a fixed formation over the earth surface. When there are more satellites in the orbit, the angle distance becomes more important. Future development of communication and space tehnology is constructing smaller earth stations with better antenna lobes. Hence the angle resolution is confined. Simply stated, the ustable frequency range is confined. In such big crush of satellites it is therefore of necessity to struggle for space of transmission capacity. 
\\
\\
In 1979 the International Telecommunication Union with headquartered in Geneva, Switzerland, tried to solve this problem by assignment of orbit positions for television satellites by giving each country a fixed orbit position. Unfortunately the assignment system errs, because the interest of national covering is a minimal advantageous from the viewpoint of telecommunication. It turns out that regional systems with a huge number of TV programs from a point of orbit are much more attractive than each orbit for each country. 
\\
\\
An other situation that complicates exploitation of the geostationary orbit is some area is more attractive than other. Since the focus of telecommunication is the populated areas, obtaining full covering of these areas is thus important. A single geostationary satellite can provide communication to large areas, as much as one-third of the earth, permitting easy interconnection between distant ground terminals. Thus three geostationary satellites placed $120^\circ$ apart can provide coverage to almost all the populated areas of the world. Hence, in the range of $10^\circ$ and $75^\circ$ east the GEO satellites will be visible over the area from Japan to Western Europe.
\\
\\
One disadvantage of geostationary satellites is that they appear almost at the horizon at latitudes above approximately $\pm 76^\circ$. Above about $\pm 84^\circ$ latitude the satellites are no longer visible. They become unusable below about $5^\circ$ elevation. This is because inadequate received signal quality caused by a combination of increased degradation in the troposhere and ground reflectios which cause rapid signal fluctuation. Thus, for providing service to high-latitude locations other types orbits like LEO and pole orbits, {\i e.g.} SvalSat, are necessary. 
\\
\\
A geostational satellite will be exposed for forces changing the parameters of the orbit. The most dominant forces in orbit perturbation are earth-west and north-south drift. In order to exploit the geostationary obital maximumthe GEO satellites must holded on by extra forces loaded in the satellites. They must be holded on within $0.1^\circ {\times} 0.1^\circ$ for maximuim exploitation. As we said, the ortibel perturbation is caued by earth-west and north-south drift. The earth-west drift is a consequence of inhomogeneities in mass distribution of the earth. It caues an east-west relocation of the satellites. They will be gathered in the stabile positions; a place south for India and the eastern Pacific. Satellites that are placed in these regions will thus be in stillness. In the Atlantic and north for New-Zealand there is a labile point such that a satellite at $19.2^\circ$ will be drawn eastwards and hence it is nessecery to induce the satellite control pulses to the west. The north-south drift is mainly caued by asymmetric influence from the sun and the moon. Hence it may be expressed as inclination drift. Today's GEO satellites have propellant on board counteracting the inclination drift which is about $1^\circ$ per year. Therefore the question about propellant is an important part of communication and space technology.
\\
\\
There are many ways to reduce the propellant requirement launching satellites to geostationary orbit. Multistage rockets are been used to reduce the propellant consumption in launching. Moreover, by utilizing multistage transfer orbit when the satellite is placed in space, will reduce the propellant requirement. In 1989 the Tele-X satellite was launched by using multistage firings of the apogee engine causing multistage raised perigee. The Tele-X satellite had made three times firings. It turns out that such a strategy requires less propellant. In this paper we will thus review the calculation of the multistage satellite launch.
\\
\\
For satellites circling around the earth, the east-west drift and north-south drift are the main effect of orbit perturbation. A special effect caused by Einstein's general relativity will perturbate an elliptic orbit, changing the orbit's perigee angle. The purpose of this paper is thus to calculate this effect as a consequence of the rotational symmetric gravitational field and discuss its application in multistage geostationary transfer orbit. This effect may ease the transition from transfer orbit to the geostationary orbit.

\section{Multistage transfer orbit} 
Let us first review a simple calculation of a one-stage transfer to the geostationary orbit. The geostationary orbit time is shorter than a solar day. During a solar day the earth rotates $360^\circ$ pluss $0.986^\circ$. Hence the geostationary orbit time reads $T = 24\times 3600 \times 365.25/364.25 {\rm s}= 86164 {\rm s}$. According to Kepler's law
\begin{equation}\label{eq:Kepler}
		T = 2\pi \sqrt{   \frac{r_m^3}{\mu}  },
\end{equation}
where $\mu = GM= 398\,603.2{\rm km}^3/{\rm s}^2$ is the product of the Newtonian gravitational constant and the mass of the earth. Hence the mean value of the orbit radius is given by \cite{Stette}
\begin{equation}
		r_m = \left[\mu \left(\frac{T}{2\pi}\right)^2 \right]^{1/3} = 42\,164 {\rm km}.
\end{equation}
The velocity follows by 
\begin{equation}
		v_{geo} =  \sqrt{\frac{\mu}{r_m} } =  3.075 {\rm km/s}.
\end{equation}
One-stage transfer launch to the geostationary orital requires a perigee point of the transfer orbit to be less $200 {\rm km}$ over the surface. When the satellite is in the apogee point, it must be supplied with a velocity increase that makes the satellite coming in geostationary orbit. Such a transfer orbit needs the paramters as appgee and perigee radius, denoted $r_a$ and $r_p$ respectively. The appgee radius must be the same as the geostationary radius, i.e. $42\,164 {\rm km}$. Considering the lowest possible perigee height at $200 {\rm km}$ over the earth surface, due to influences from the atmosphere, the perigee radius becomes $6\,578 {\rm km}$. The mean value of these distances is obvious $24371 {\rm km}$. According to Kepler's law (\ref{eq:Kepler}), the transfer orbit time is calcuted to be $37\,864 {\rm s}$. Hence, by using Newton's law of gravity, $\mbox{d}^2{\bf r}/\mbox{d}t^2 = -G M{\bf r}/r^3$, the apogee velocity is given by
\begin{equation}
		v_{a} =  \sqrt{ 98603.20 \left( \frac{2}{42164}  - \frac{1}{24371} \right)  }  {\rm km/s} = 1.597 {\rm km/s},
\end{equation}
which yields a required speed increase $\Delta v=  1.496 {\rm km/s}$. This is obvious a very high velocity increase at the apogee point of the transfer orbit requiring high propellant loaded in the satellite. However, the velocity increase is valid only if the transfer orbit is situated in the same plane as the geostationary orbit. If not, there is an inclination compared with the equatorial plane that must be eliminated resulting in a higher value than $1.597 {\rm km/s}$. The requirement of propellant is very high in a such one-stage transfer launch. The rocket equation derived from Newton's second law can tell us what magnitude of propellant the satellite needs. The rocket equation reads
\begin{equation}
\label{eq:rocket}
		\Delta  v = v_p \int_{m_0 + m_p}^{m_0} {\mbox{d}m/ m} = v_p \ln\left(\frac{m_p + m_0 }{m_0}\right)
\end{equation}
where $m_0$ is the mass of the satellite. $m_p$ and $v_p$ are the mass and the velocity of the total propellant. According to Eq. (\ref{eq:rocket}), the mass of the propellant is given by 
\begin{equation}
\label{eq:rocketRatio}
		m_p = m_0 \left[\exp\left( \frac{\Delta v}{v_p}\right)-1\right],
\end{equation}
where $v_p$ depends on which type of propellant we have selected. If we consider one-fire-stage solid engine which gives $v_p= 3 {\rm km/s}$, the mass of propellant must be less $63.26\%$ of $m_0$ stored in the satellite. It is enormous unapplicable and unprofitable in space technology. However, using ion engine developed in 2002 will reduce the propellant considerably. An ion engine has a very high velocity of flow making it possible to reduce the mass of propellant. A typical ion enegine may provide $v_p = 30 {\rm km/s}$. Consequently, the propellant is reduce to $5\%$ which is very good. Ion engines have small push power making them only applicable for slow controls in space. The propellant in ion engines has been ionized xeon gas. These engines are ten times more effective than chemical rocket engines. Moreover, ion enegines provide longer lifetime for the satellite. These engines will be one of the most important space engines in the future.
\\
\\
Although we can reduce the satellite propellant by choosing ion engine build on the satellite, the total required propellant is still high to launch a rocket to the geostationary orbit. Hence it is not enough to improve propellant and crossing the most suitable propellant and engine, we must also improve our strategy for launching the rocket and placing the satellite from a transfer orbit to the geostationary orbit. 
\\
\\
Apogee engines with solid propellant cannot be stopped and restarted. They suitable burn only once. With liquid propellant, the engine can be turn on and off several times. However, engines with liquid propellant are very difficult to contruct. 
\\
\\
Multistage rokcets is prefered when we consider a launch from the earth surface. Such rockets reduce the propellant a lot. Simple calculation of the rocket equation indicates that. Assuming that the mass of the rocket without propellant is given by the expression
\begin{equation}
\label{eq:rocketStage1}
		m_r = K\left(  m_p + m_s   \right),
\end{equation} 
where $K$ is an arbitrary constant and $m_s$ is the mass of payload. The expression (\ref{eq:rocketStage1}) is reasonable since the mass of the rocket must be larger than the sum of payload and propellant. Applying Eq. (\ref{eq:rocketStage1}) on the rocket equation (\ref{eq:rocket}) written in exponential form, we obtain
\begin{equation*}
		\exp\left(\frac{\Delta v}{v_p}\right) = \frac{(m_p + m_s)(K+1)}{K(m_s + m_p)+ m_s},
\end{equation*} 
which gives the ratio of propellant and payload
\begin{equation}
\label{eq:rocketStage2}
		\frac{m_p}{m_s}  = \frac{\exp\left(\frac{\Delta v}{v_p}\right) -1}{ 1- \frac{K}{K+1}\exp\left(\frac{\Delta v}{v_p}\right) }.
\end{equation} 
The ratio goes to infinity when the denominator goes to zero, i.e. $\left[K/(K+1)\right]\exp\left(\Delta v/v_p\right)=1$. The maximum velocity increase for a finial payload will be lower when the ratio of $K$ and $K+1$ goes to infinity. Realistically the increase is set to be $v_p=3 {\rm km/s}$. The ratio of the rocket  and the sum of propellant and payload is typical set to be $K= 0.1$. Thus, it gives $\Delta v = 7.2 {\rm km/s}$, which is the highest velocity a one-stage rocket can obtain. As we see, it is not so impressive. Three-stage rocket is then suggesed and it seems to be very good. To obtain a total velocity at $10.5 {\rm km/s}$ using three-stage rocket, we only need a velocity increase for each stage requires $3.5 {\rm km/s}$. In this case the third stage propellant is calculated to be $m_{p}^{(3)}= 3.12 \, m_s$, when we assume $K= 0.1$ and $v_p= 3 {\rm km/s}$. The total mass expressed by $m_s$ is calculated to be $(m_s + 3.12\,m_s +0.1\cdot 3.12\,m_s)= 4.53\, m_s$. It is acceptable regarding construction of the rocket and the propellant quantity. Here, $1/4.53$ part of the start mass is payload, i.e satellite.
\\
\\
When it comes to getting a satellite from a transfer orbit to a geostationary orbit, we have some strategies providing reduced propellant. By applying several separate firings of the apogee engine obtaining the so-called multistage transfer orbit, the perigee point of the transfer orbit lifts each time the apogee engine fires. This strategy is applicable and favourable considering the restriction of propellant in a satellite. Let us then calculate the propellant requirement of a three-stage transfer. 
\\
\\
Assuming the perigee is $200 {\rm km}$ over the earth surface for the first transfer orbit. Then the second perigee becomes $11\,862 {\rm km}$ higher than the first one, and so on until we reach the radius of the geostationary orbit. At the first transfer orbit, the apogee velocity is calculated to be $1.597 {\rm km}$. Firing the apogee engine to increase the perigee with $11862 {\rm km}$, the apogee velocity increase $\Delta v_1= 0.454 {\rm km/s}$ is required. The rocket equation gives the following expression of the mass of propellant needed to go from the first transfer orbit to the next one,
\begin{equation}
    m_p^{(1)} = m_0 \left(\xi_1 - 1\right) + \left[m_p^{(2)}+ m_p^{(3)}\right]\left(\xi_1 - 1\right),
\end{equation} 
where $\xi_1 = \exp \left(\Delta v_1/v_p\right)$. $m_p^1$, $m_p^2$ and $m_p^3$ are the propellant mass at the first, second and third stage respectively. The next stage is to reach the perigee $r_p = 23924 {\rm km}$ which requires an apogee velocity at $2.6162 {\rm km/s}$. Hence the velocity increase is $\Delta v_2= 0.5662 {\rm km/s}$ producing 
\begin{equation}
    m_p^{(2)} = m_0 \left(\xi_2 - 1\right) +  m_p^{(3)} \left(\xi_2 - 1\right)- m_p^{(1)},
\end{equation} 
where $\xi_2 = \exp \left(\Delta v_2/v_p\right)$. The last transfer is to increase the perigee equal to the apogee, i.e. the radius of the geostationary orbit. Consequently, the last apogee velocity must be $v_{geo}=3.075 {\rm km}$ resulting in $\Delta v_3= 0.4588 {\rm km/s}$. According to the rocket equation we obtain the following expression 
\begin{equation}
	\xi_3  = \exp \left(  \frac{\Delta v_3}{v_p}  \right) = 1- \frac{ m_p^{(1)} +  m_p^{(2)}  +  m_p^{(3)}}{m_0}.	
\end{equation} 
Considering solid propellant in the satellite, i.e. $v_p=3 {\rm km/s}$, we obtain $\xi_1 = 1.1633$, $\xi_2 = 1.2077$ and $\xi_3 = 1.16523$. The last one is the important to calculate the total required propellant in the satellite $m_p = m_p^{(1)} +  m_p^{(2)}  +  m_p^{(3)}$. Obviously we see that the propellant needed is $16.53\%$ of the mass of the satellite. With such a simple calculation, we see that we may reduce the propellant by applying multistage rocket and multistage transfer orbit. This is very much less than one-stage transfer orbit strategy which requires $m_p=63.18\%\,m_0$. It is therefore not inconsequent that the Tele-X satellite launch in 1989 had using this three-stage transfer strategy.

\section{Effect of the General Theory of Relativity} 
As we have seen in previous section that we can reduce the propellant in the satellite by using multistage transfer orbit. However, it should be a great benefit if we can utilize some orbit perturbation to reduce the apogee velocity increase, i.e reduce the propellant. Early in the 20th century, it is observed that the perigee point of the orbit of Mercury relocates a small angle for each revolution. Hence it is interesting to give a review of rotational symmetric gravitational field and how it affects the transfer orbit of geostationary orbit launch.
\\
\\
General relativity is a theory of gravitation derived by Albert Einstein in 1915. It is an amplification of the Newtonian theory of gravitation. The theory states an equality between gravity and curved spacetime. Einstein's theory of gravity is a second order differential equation of the so-called metric tensor $g_{\mu \nu}$, where second derivative comes in through the tensor of curvature. Hence the Einstein equation of gravity gives a direct connection between the curvature of the four-dimensional spacetime and the total energy-momentum-tensor of all particles and all fields except the field of gravitation itself. Without detailed derivation, the Einstein equation of gravity is given by Einstein's tensor of curvature
\begin{equation}
	G_{\mu \nu} = \kappa \,T_{\mu\nu},
\end{equation} 
where $\mu, \nu = 0,1,2,3$. The index $0$ denotes the time of the tensor. Einstein's gravitational constant is denoted by $\kappa$ which is proportional to Newton's constant $G$, $\kappa= 8\pi G/c^4$. $T_{\mu\nu}$ is the energy-momentum tensor. The trace of Einstein's tensor of curvature is the scalar curvature denoted by $R$, reading $G={\rm Tr}\, G_{\mu \nu}= g^{\mu \nu }G_{\mu \nu} = g^{\mu \nu }R_{\mu \nu} - 4R/2 = -R$. It results in $R= -\kappa T$, where $T= {\rm Tr} \,T_{\mu \nu}= g^{\mu \nu }T_{\mu \nu}$, providing an other equivalent form of the Einstein equation
\begin{equation}
	R_{\mu \nu} = \kappa\, \left( T_{\mu\nu} -\frac{1}{2}g_{\mu \nu}T\right).
\end{equation} 
After the Einstein equation became well-known in 1915, Schwarzschild came up with an exact solution of the equation. The solution is found by considering a rotational symmetric gravitational field with spherical symmetric metric. The solution is called the Schwarzschild metric 
\begin{equation}
	\mbox{d}s^2 = \left( 1- \frac{2GM}{c^2r}\right) c^2\mbox{d}t^2 - \mbox{d}r^2 \frac{1}{1-\frac{2GM}{c^2r}}- r^2 \left(  \mbox{d}\theta^2   + \sin ^2\theta \mbox{d}\varphi^2 \right),
\end{equation} 
where $t$ is the time and $M$ is the gravitational mass, i.e. the mass of the earth in  this case, which is constant. The solution describes the gravitational field in vacuum outside an arbitrary spherical mass distribution with a total mass $M$. Inside the mass discontribution we have to modify the metric. Outside we are free to choose the space and the time coordinates. 
\\
\\
Hence, the Schwarzschild metric must be a good approximation of the gravitational field from any approximate spherical bodies. In the sense of satellite orbits, the metric is good as well for the gravitational field from the earth considering not too large distant and neglectable contribution from the moon and the other planets. Hence, the equations of motion can be carried out by the principle of variation, i.e extremalizing the action integral $S= -mc\int{w\mbox{d}u  }$, where $m$ is the mass of the satellite and $u$ is an arbitrary parameter along the path, i.e. the orbit. Thus we can define $w= \mbox{d}s/\mbox{d}u$. Applying $w$ in the Euler-Lagrange equations and doing some algebra gymnastics, it will produce a second order differentional equation 
\begin{equation}
\label{eq:GReffect}
		\frac{\mbox{d}^2\psi}{\mbox{d}\varphi^2} = \frac{1}{2B^2}- \psi + \frac{3}{2}\psi^2,
\end{equation} 
where $\psi = 2GM/c^2r$ and $B$ is a dimensionless integration constant proportional to the momentum of the $z$-axis $L_z$, which is canonical conjugate momentum of the azimuth angle $\varphi$. The angle corresponds to the relocation of the elliptic transfer orbit. The last term in Eq. (\ref{eq:GReffect}) is the pure effect of the general relativity. The rest agrees with the classical Newtonian gravitational law. Rewriting Eq. (\ref{eq:GReffect}) as a function of the angle, we obtain the expression of the movement of the perigee point of an elliptic orbit
\begin{equation}
\label{eq:GReffect1}
		\Delta \varphi = \frac{6\pi GM}{c^2r_0},
\end{equation}
for each circulation where $r_p = r_0/(1+e)$ and $r_a = r_0/(1-e)$. For $r_p = 6578{\rm km}$, i.e. $200{\rm km}$ over the earth surface, and $r_a = 42164{\rm km}$, the eccentricity $e$ and the middle distance $r_0$ are found to be $0.73$ and $11380.525{\rm km}$. Hence, the mean distanse $r_m = 2.1414653 \,r_0$. Inserting $M$ equal to the mass of the earth $5.974\times 10 ^{24}{\rm kg}$, Eq. (\ref{eq:GReffect1}) gives a change of the orbit with the angle $\Delta \varphi = 7.34655 \times 10^{-9}$, i.e. $1.58686 \times 10^{-3}"$ per circulation around the earth. A three-stage transfer orbit with a perigee $200 {\rm km}$ over the surface, the gravitational effect gives a change of the orbit corresponding to $\Delta \varphi = 4.74757 \times 10^{-3}"$. Although the change is neglictable and not discernible, it move the apogee point of the transfer orbit almost one meter, $0.967 {\rm m}$. This effect reduces the required apogee velocity increase with $\left(0.967 /86164\right){\rm m/s} = 1.0756 \times 10 ^{-5} {\rm m/s}$. Accordingly, the gravitational effect does not affect the velocity increase $1.496 {\rm km /s}$ which is needed for getting the satellite from a $200{\rm m}$ perigee transfer orbit to GEO.  According to Eq. (\ref{eq:rocketRatio}) and considering solid propellant, this effect reduce the propellant with $\Delta m_p= 3.585\times 10^{-9}\, m_0$ which is very neglectable. The reduced propellant is also neglectable using ion engines. $1.0756 \times 10 ^{-5} {\rm m/s}$ is also to small to give any affect to a three-stage transfer orbit, where $\Delta v = 458.8{\rm km/s}$.
\\
\\
Although the effect of general relativity is neglectable considering utilization of getting satellites from a transfer orbit to GEO. However, it affects the satellite platform and the directional properties since the effect causes one meter movement of the apogee point for each circulation for the GEO transfer orbit. The effect is important when it comes to highly elliptical satellite orbits, {\i e.g.} Molniya orbit. Several elliptical circulations around the earth will also increase this effect.

\section{Discussion on space environment} 
Being in orbit around the earth is not unproblematic. High energy particles situated in the atmosphere can cause damages on the satellites circulating around the earth. It mainly regards free electrons and protons from the space and the sun. Flux and energy of these particles are determined by height over the earth surface and the solar activity. However, galactic cosmic rays constitute $90\%$ of the protons in addition to some alpha-particles with the energy in the range GeV. The flux of those particles is about $2.5$ per ${\rm cm}^2/{\rm s}$. High energy particles can damage to electronic in the satellites. If a mechanical collision with lattice structure occurs in a semiconductor, it may cause errors leading to changed electrical properties. In other hands, if positive or negative particles penetrate into a lattice structure, the materials will be ionized resulting in damages to components made of semiconductors and to dielectric materials in the satellites. Hence, the long term effect of radiation requires good protection and technical costly.
\\
\\
Launching satellites to goestationary orbit using multistage transfer orbit with low perigee is extra complicated due to the van Allen belts. van Allen belts are a zone surrounding the earth containing charged particles captured by the magnetic field of the earth. The composition of particles varyes with the height over the earth surface. The van Allen belt consists in two types, the external belt consisting electrons and the internal belt consisting protons. The protons reside mostly within $25,000{\rm km}$ height, with its maximum flux $2\times 10^{6}{\rm cm}^{-2}{\rm s}^{-1} $ at $11,000{\rm km}$. However, the crossing is not abrupt.
\\
\\
The radiations of protons and electrons will damage a satellite residing in the van Allen belts for a long time. It requires very advanced protection resulting in unprofitable utilization of multistage transfer orbital with low perigee. To avoid damages caused by the van Allen belts we must lift the perigee over the critical height $200-300{\rm km}$. Hence, it is important to choose transfer orbits with their perigee over this height. It causes less elliptical GEO transfer orbit, i.e. less effect of general relativity. 
\\
\\
Moreover, there are many other gravitational effects that we do not have counted on. These forces are caused by heavenly bodies. For low earth satellites the effects of gravitational forces of gravitational forces from the sun and the moon are small compared with the gravitational force of the earth. However, these causes noticeable perturbations to a geostationary orbit. The main sources of perturbations on a geostationary satellite, in order of magnitude, are the gravitational forces of the sun and the moon. The satellite in the GEO orbit receives a stronger gravitational pull in the direction of the heavenly bodies when nearer to them, causing a gravity gradient. The main effect of this gradient is changing the inclination of the satellite orbit. The combined effect of the sun and the moon causes a change in the inclination of a geostationary satellite $0.75^\circ - 0.94^\circ$. The change in the inclination of the orbit caused by the sun is about $0.27^\circ$ per year. The change is steady of the accuracy considered here as the relative position between the satellite and the ecliptic plan of the earth around the sun remains almost fixed. Generally, there are three forces affecting the inclination of a satellite. The first one is the gravitational pull of the sun and the moon. Then a force due to the non-spherical nature of the earth. The latter one has a component in a direction opposite to the former two forces, canceling out at an inclination about $7.5^\circ$. Accordingly, the inclination of a geostationary satellite, oscillates around this stable inclination with a period of about $53$ years, reaching between $0^\circ$ and $15^\circ$.

\section{CONCLUDING REMARKS}\label{conclusion}
In this paper we have given a review of launching a satellite to the geostationary orbit using multistage transfer orbit. Moreover we have calculated the effect of the rotational symmetric gravitational field. Although this effect is neglectable, it does affect the apogee velocity increase and the position of the apogee of a GEO transfer orbit. Several elliptical circulations will increase the effect and hence reduce the propellant need to apogee velocity increase. Unfortunately, with today's technology, the cost of plenty circulations will be larger than the reduced propellant cost regarding potential damages of the satellite near the van Allen belts. 
\\
\\
Before getting too carried away, it is certainly clear that we are far from profitable utilization of the gravitational effect to reduce the propellant. As of today, any space applications does not take this effect into account. However, if technology allows satellites situating in an elliptical transfer orbit for a very long time, the effect of general relativity may be of benefit in space traveling in future.

\section{ACKNOWLEDGMENTS}  
This paper is carried out at the Department of Physics at Norwegian University of Science and Technology (NTNU) as an indenpendent research and as a problem given in the couse FY3120. The paper has been worked up during the fall of 2005.

\end{document}